# THE P-VALUE FROM A FUZZY POINT OF VIEW

**Piero Quatto**
*Department of Economics, Management and Statistics, University of Milano – Bicocca, Milan, Italy*
E-MAIL: piero.quatto@unimib.it
ORCID: 0000-0002-6679-7169

# THE P-VALUE FROM A FUZZY POINT OF VIEW

***Abstract.*** *The purpose of the paper is to provide a new way of seeing the p-value in terms of a fuzzy membership function. According to the ASA's statement, we aim at removing the arbitrary choice of the significance level and at demonstrating that the p-value can be profitably interpreted from a fuzzy point of view. In particular, we propose a new class of membership functions by viewing the p-value as a function of the null hypothesis and we apply our approach to compare two independent binomial proportions. The proposed membership functions can also be employed to assess the precision of confidence intervals and the power of statistical tests.*

***Keywords:*** *P-value, Confidence intervals, Fuzzy sets, Comparing binomial proportions.*

## 1. INTRODUCTION

The logic of significance tests can be explained entirely in terms of a well-known disjunction (Fisher, 1956): if it follows from the null hypothesis $H_0$ that an observed event is very improbable, then either a very rare event has occurred or the null hypothesis is very unlikely and so it should be rejected.

According to this logic, tests of significance are capable of rejecting $H_0$, in so far as the null hypothesis is contradicted by the data, but they are never capable of ascertaining $H_0$.

In order to give a measure of how strongly the observations contradict the null hypothesis, we can provide the p-value of the test, defined as the probability under $H_0$ of getting results as extreme as those observed.

Although these arguments seem plausible, it is not known where to draw the line between unlikely and likely hypotheses, that is to say, it is not clear how to select the threshold $\alpha$ between significant and non-significant p-values, easily generating misinterpretations (Greenland, 2016).

Referring to such a fundamental question, the American Statistical Association (ASA) has released an important statement on the use of p-values, which consists of the following six principles (Wasserstein and Lazar, 2016):

- p-values can indicate how incompatible the data are with a specified statistical model;
- p-values do not measure the probability that the studied hypothesis is true, or the probability that the data were produced by random chance alone;
- scientific conclusions and business or policy decisions should not be based only on whether a p-value passes a specific threshold;
- proper inference requires full reporting and transparency;
- a p-value, or statistical significance, does not measure the size of an effect or the importance of a result;
- by itself, a p-value does not provide a good measure of evidence regarding a model or hypothesis.

In agreement with all the six ASA's principles, our paper aims at removing the arbitrary choice of the significance level $\alpha$ and at demonstrating that p-values offer useful information, which can be easily interpreted from a fuzzy point of view.

For these purposes, in Section 2 we introduce the motivating example, which is used throughout the paper and relies on comparing two independent binomial proportions; in Section 3 we present a new class of fuzzy membership functions by viewing the p-value as a function of the null hypothesis; finally we draw conclusions in Section 4.

## 2. MOTIVATING EXAMPLE

The comparison of two treatments is a problem arising in many different contexts where the observations consist of the number of successes in a sequence of independent trials for each treatment (Mehrotra et al., 2003), and so the natural model is provided by two independent binomial variables
$$X \sim Bin(m, \omega)$$
and
$$Y \sim Bin(n, \omega + \theta)$$
with four parameters representing the two sample sizes ($m$ and $n$) and the two probabilities of success ($\omega$ and $\omega + \theta$).

In particular, we consider a null hypothesis on the parameter of interest $\theta$ and the test statistic

$$\left|\frac{Y}{n} - \frac{X}{m} - \theta\right|$$

measuring the distance between the difference of proportions and the corresponding difference of probabilities specified by $\theta$.

We can therefore calculate the probability of getting results as extreme as those observed by means of

$$P_{\theta,\omega}\left(\left|\frac{Y}{n} - \frac{X}{m} - \theta\right| \geq \left|\frac{y}{n} - \frac{x}{m} - \theta\right|\right)$$
$$= \sum_{(u,v) \in C} \binom{m}{u} \omega^u (1-\omega)^{m-u} \binom{n}{v} (\omega+\theta)^v (1-\omega-\theta)^{n-v},$$

where $x$ and $y$ are the observed values of the random variables $X$ and $Y$, respectively, $\omega$ is a nuisance parameter with range given by

$$max\{0, -\theta\} < \omega < min\{1, 1-\theta\}$$

with

$$-1 < \theta < 1,$$

and

$$C = \left\{(u,v) \in \mathbb{N} \times \mathbb{N} : u \leq m, v \leq n, \left|\frac{v}{n} - \frac{u}{m} - \theta\right| \geq \left|\frac{y}{n} - \frac{x}{m} - \theta\right|\right\}.$$

The choice of the motivating problem is also due to the presence of a nuisance parameter, which allows us to better illustrate the potential of the approach developed in Section 3.

The p-value by itself does not offer a consistent general-purpose measure of evidence and hence scientific conclusions should not be based only on whether or not a p-value is below some fixed level $\alpha$ (usually set at 0.05), as stated by the ASA's principles (Wasserstein and Lazar, 2016).

Since a single p-value alone provides very limited information, we propose to supplement this classical statistic in a fuzzy perspective.

To give effect to this purpose we consider the following case study (Villiger et al., 2016).

The study is a single centre, phase 2, randomised, double-blind, placebo-controlled trial, aimed to show the efficacy of Tocilizumab, a humanized monoclonal antibody against the interleukin-6 receptor, in rapid induction and maintenance of remission in patients with giant cell arteritis. Thirty patients aged 50 years and older who fulfilled the 1990 American College of Rheumatology criteria for giant cell arteritis were recruited at the University Hospital Bern in Switzerland between March 2012 and September 2014. Patients were randomly assigned to receive glucocorticoids and either Tocilizumab at 8 mg/kg bodyweight (20 patients) or placebo (10 patients), both intravenously. Patients received 13 infusion every 4 weeks until week 52. The primary endpoint was complete remission at week 12 without clinical sign or symptoms. Specifically, $y = 17$ of $n = 20$ patients given Tocilizumab (85%) and $x = 4$ of $m = 10$ patients given placebo (40%) reached complete remission by week 12.

The choice of the case study is also due to the small sample sizes, which do not allow to oversimplify the inferential problem by resorting to asymptotic approximations.

## 3. MEMBERSHIP FUNCTIONS

In this section we show that p-values can be seen from a fuzzy point of view as membership functions.

### 3.1 FUZZY SETS

A fuzzy set $A$ in a universe of discourse $U$ is defined by means of a membership function
$$\mu_A: U \to [0,1]$$
yielding the membership grade $\mu_A(u)$ of each element $u \in U$ to the fuzzy set $A$. In particular, $A$ is included in the fuzzy set $B$ if
$$\mu_A(u) \leq \mu_B(u) \quad \forall u \in U$$
and the inclusion is denoted by
$$A \leq B$$

(Bede, 2013).

Following Dutta et al. (2011), we define the height of a fuzzy set as the largest membership grade

$$\sup_{u \in U} \mu_A(u)$$

and a strong $\alpha$-cut as the subset of $U$

$$\{u \in U : \mu_A(u) > \alpha\}$$

with $\alpha \in (0,1)$.

For instance, we can model the linguistic expression "about 10" by using a fuzzy set $A$ (in the universe of discourse $U = \mathbb{R}$), for which we may define the membership function

$$\mu_A(u) = \begin{cases} 1 - |u - 10| : & 9 \leq u \leq 11 \\ 0 & : u < 9 \text{ or } u > 11 \end{cases}$$

and so the height of $A$ and the strong $\alpha$-cut are given by

$$\sup_{u \in \mathbb{R}} \mu_A(u) = 1$$

and

$$\{u \epsilon \mathbb{R} : \mu_A(u) > \alpha\} = \,]9 + \alpha, 11 - \alpha[$$

respectively. We note, however, that the expression "about 10" can also be modelled by the fuzzy set $B$ with membership function

$$\mu_B(u) = \begin{cases} 1 - \dfrac{|u - 10|}{2} : & 8 \leq u \leq 12 \\ 0 & : u < 8 \text{ or } u > 12 \end{cases}$$

and hence we obtain

$$A \leq B$$

as shown in Figure 1.

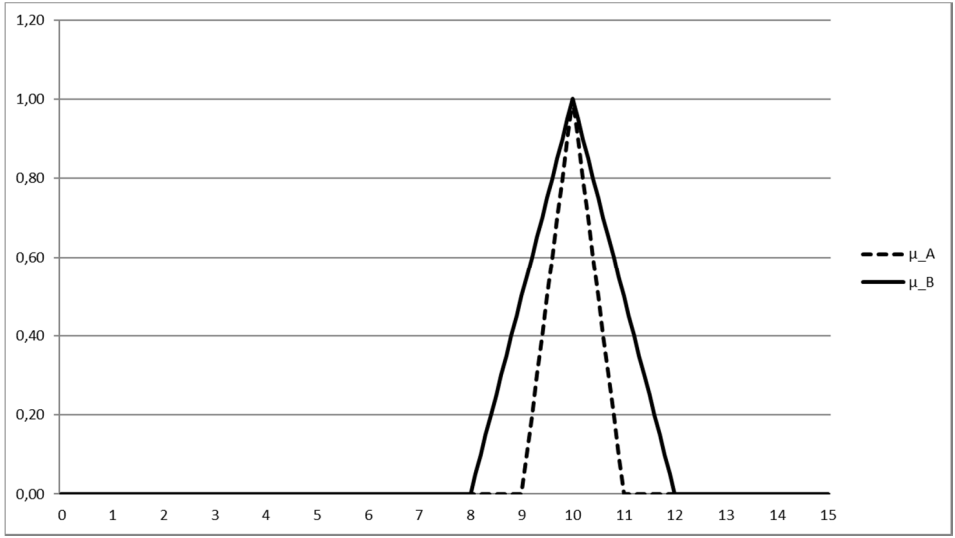

**Figure 1: Two membership functions for modelling the expression "about 10"**

### 3.2 MEMBERSHIP FUNCTIONS BASED ON P-VALUES

In order to introduce the fuzzy set $\Theta$ consisting of all plausible values taken on by a parameter $\theta$ (lying in an appropriate parameter space), we consider

- $n$ independent random variables $X_i$ ($i = 1, \ldots, n$) with the same cumulative distribution function
$$F(x; \theta, \omega) = P_{\theta,\omega}(X_i \leq x),$$
where $\theta$ is the parameter of interest and $\omega$ represents a nuisance parameter (or, for the sake of generality, a vector of parameters);
- a suitable test statistic
$$T_{\theta_0} = t(X_1, \ldots, X_n; \theta_0)$$
such that large values of the statistic calculated on the observed values $x_1, \ldots, x_n$ give evidence against the null hypothesis $\theta = \theta_0$;
- and, finally, the corresponding p-value
$$p(x_1, \ldots, x_n; \theta_0) = \sup_{\omega} P_{\theta_0, \omega}[T_{\theta_0} \geq t(x_1, \ldots, x_n; \theta_0)].$$

So, we can extend the p-value to a general hypothesis $H_0: \theta \in \Theta_0$ by means of

$$p(x_1, \ldots, x_n; \Theta_0) = \sup_{\theta \in \Theta_0} p(x_1, \ldots, x_n; \theta).$$

More precisely, a p-value $q(x_1, \ldots, x_n)$ for $H_0: \theta \in \Theta_0$ is called valid if
$$P_{\theta,\omega}[q(X_1, \ldots, X_n) \leq \alpha] \leq \alpha \quad \forall \theta \in \Theta_0, \forall \omega, \forall \alpha \in \,]0,1[$$
(Casella and Berger, 2002) and, on the basis of this definition, the following two theorems can be proved (for convenience, proofs are collected in the Appendix).

**Theorem 1**
  a) The inequality
$$P_{\theta,\omega}[p(X_1, \ldots, X_n; \theta) \leq \alpha] \leq \alpha$$
  holds for all $\theta$, $\omega$ and $\alpha \in \,]0,1[$.
  b) The p-value $p(x_1, \ldots, x_n; \Theta_0)$ associated with the general hypothesis $H_0: \theta \in \Theta_0$ is valid.
  c) The set
$$I_{1-\alpha}(x_1, \ldots, x_n) = \{\theta: p(x_1, \ldots, x_n, \theta) > \alpha\}$$
  represents a confidence interval of level $1 - \alpha$ for the parameter of interest.

Once the data $x_1, \ldots, x_n$ have been observed, we can define the membership function of the fuzzy set $\Theta$ by viewing the p-value as a function of the parameter $\theta$
$$\mu_\Theta(\theta) = p(x_1, \ldots, x_n; \theta),$$
that is to say, by measuring the compatibility of any $\theta$ value (as a part of the parametric model in question) with the observations.

According to the Theorem 1, such a function allows us to simultaneously glance at interval estimation of level $1 - \alpha$ through the strong $\alpha$-cut
$$I_{1-\alpha} = \{\theta: \mu_\Theta(\theta) > \alpha\}$$
(in analogy with Buckley, 2006) and at statistical testing of $H_0: \theta \in \Theta_0$ via a valid p-value that can be interpreted as the height restricted to the set $\Theta_0$
$$p(\Theta_0) = \sup_{\theta \in \Theta_0} \mu_\Theta(\theta).$$

As required by the ASA's statement (Wasserstein and Lazar, 2016), this approach permits full reporting and transparency of statistical analyses, since it enables the researcher to disclose all hypotheses explored and all p-values calculated by resorting to a graphical representation, which is more informative that any single test and is similar to the confidence curves suggested by Birnbaum (1961) and Bender et al. (2005).

In particular, the proposed membership function avoids the need of fixing an arbitrary significance or confidence level.

By means of such membership functions, we can also evaluate the precision of confidence intervals and, consequently, the power of statistical tests. In practice, for comparing two statistical tests (on the basis of the same observations and for the same parameter $\theta$), leading respectively to the membership functions $\mu_{\Theta_1}$ and $\mu_{\Theta_2}$, we shall prefer the first test to the second one if

$$\Theta_1 \leq \Theta_2,$$

that is to say, if $\mu_{\Theta_1}$ provides confidence intervals included in those (of the same level) supplied by $\mu_{\Theta_2}$.

For the motivating example introduced in Section 2, the fuzzy set of all plausible values taken on by $\theta$ can be defined in terms of the membership function

$$\mu(\theta) = \sup_{\omega} P_{\theta,\omega}\left(\left|\frac{Y}{n} - \frac{X}{m} - \theta\right| \geq \left|\frac{y}{n} - \frac{x}{m} - \theta\right|\right),$$

where $-1 < \theta < 1$ and the supremum is taken over all $\omega$ values between $max\{0, -\theta\}$ and $min\{1, 1 - \theta\}$. Figure 2 shows how such a membership function can be used effectively in interval estimation and in statistical testing at every possible level. For instance, the horizontal dashed line represents a confidence interval of level 0.95 and the vertical solid line represents the p-value for $H_0: 0 \leq \theta \leq 0.2$ (horizontal solid line) given by

$$\sup_{0 \leq \theta \leq 0.2} \mu(\theta) = 0.0236.$$

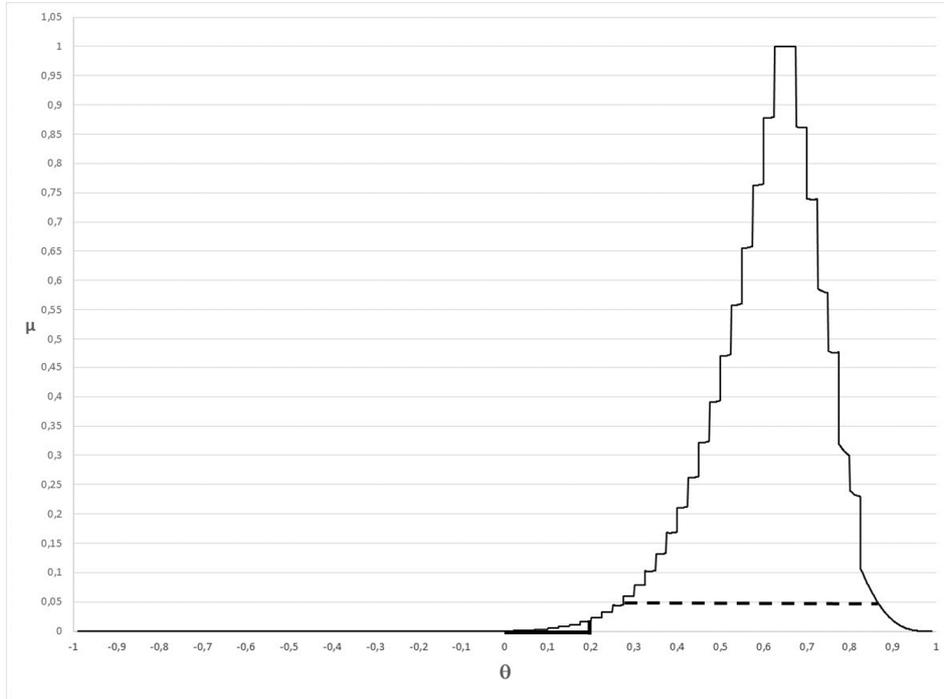

**Figure 2: The membership function for the motivating example**

Besides, we can employ all the available information concerning the nuisance parameter to improve the Theorem 1 in terms of power (Berger and Boos, 1994) by introducing

$$b(x_1, \ldots, x_n; \theta) = \gamma + \sup_{\omega \in S} P_{\theta, \omega}[T_\theta \geq t(x_1, \ldots, x_n; \theta)],$$

where $S = S(x_1, \ldots, x_n)$ is a confidence set of level $1 - \gamma$ for the nuisance parameter, that is

$$P_{\theta, \omega}[\omega \in S(X_1, \ldots, X_n)] \geq 1 - \gamma \quad \forall \theta \; \forall \omega.$$

**Theorem 2**
a) $P_{\theta, \omega}[b(X_1, \ldots, X_n; \theta) \leq \alpha] \leq \alpha \quad \forall \theta, \forall \omega, \forall \alpha \in \,]0,1[.$

b) The p-value of $H_0: \theta \in \Theta_0$
$$b(x_1, \ldots, x_n; \Theta_0) = \sup_{\theta \in \Theta_0} b(x_1, \ldots, x_n; \theta)$$
is valid.
c) The set
$$B_{1-\alpha}(x_1, \ldots, x_n) = \{\theta: b(x_1, \ldots, x_n, \theta) > \alpha\}$$
is a confidence interval of level $1 - \alpha$.

Returning to the motivating example of Section 2 and according to Theorem 2, we can improve the membership function by using
$$\mu^S(\theta) = \gamma + \sup_{\omega \in S} P_{\theta,\omega}\left(\left|\frac{Y}{n} - \frac{X}{m} - \theta\right| \geq \left|\frac{y}{n} - \frac{x}{m} - \theta\right|\right),$$
where $-1 < \theta < 1$ and the set
$$S = \left\{\omega \in \left]\frac{x}{m} - z\sqrt{\frac{x(m-x)}{m^3}}, \frac{x}{m} + z\sqrt{\frac{x(m-x)}{m^3}}\right[ : max\{0, -\theta\} < \omega \right.$$
$$\left. < min\{1, 1 - \theta\}\right\}$$

is based on an approximate confidence interval of level $1 - \gamma$ for $\omega$, with $z$ denoting the $1 - \gamma/2$ quantile of the standard Normal distribution (Agresti and Coull, 1998). Figure 3 shows that for $\gamma = 0.0001$
$$\mu^S(\theta) \leq \mu(\theta) \quad \forall \theta \in \,]0,1[$$
but the improvement does not seem to be substantial.

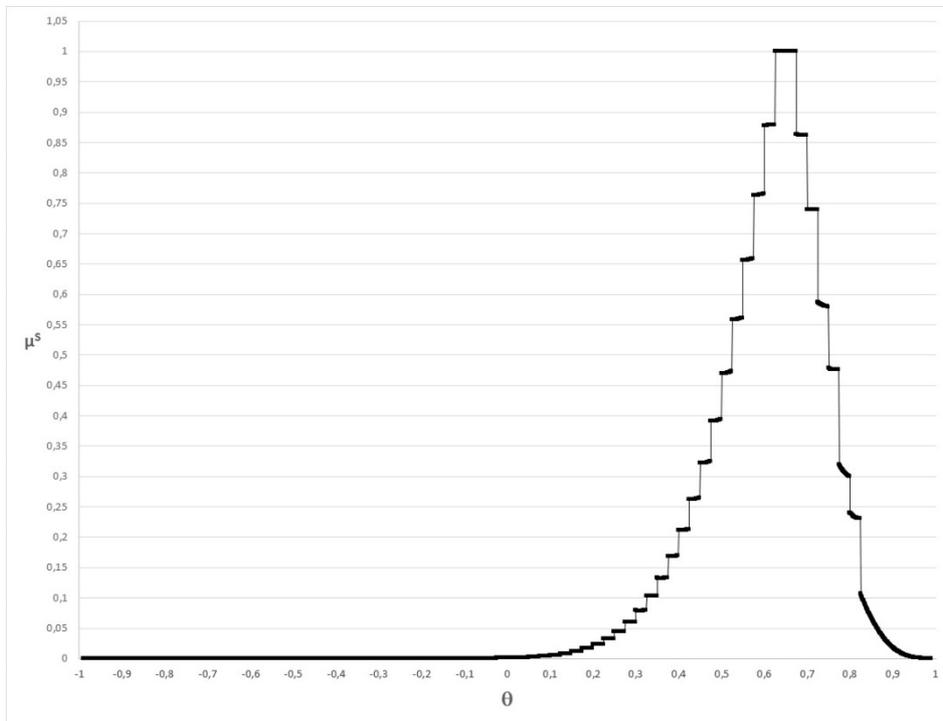

**Figure 3: The membership function for the motivating example based on Berger and Boos (1994)**

## 4. CONCLUSIONS

We presented as the inference a membership function on the set of possible values for the quantity of interest, where the membership function is dependent on the data through the p-value.

Membership functions based on p-values can be used effectively in interval estimation and in statistical testing at all possible levels, avoiding the need of fixing an arbitrary confidence or significance level, and resulting in graphical representations, which are more informative that any single interval or test, as shown by the motivating example.

In addition, such membership functions can be employed to assess the precision of confidence intervals and the power of statistical tests.

**APPENDIX**

*Proof of Theorem 1.*
   a) By defining the cumulative distribution function
   $$G(y; \theta, \omega) = P_{\theta,\omega}(Y \leq y)$$
   of the auxiliary statistic $Y = -T_\theta$, we obtain
   $$p(x_1, \ldots, x_n; \theta) = \sup P_{\theta,\omega}[T_\theta \geq t(x_1, \ldots, x_n; \theta)]$$
   $$\geq P_{\theta,\omega}[T_\theta \geq t(x_1, \ldots, x_n; \theta)] = G(-t(x_1, \ldots, x_n; \theta); \theta, \omega)$$
   and then
   $$P_{\theta,\omega}[p(X_1, \ldots, X_n; \theta) \leq \alpha] \leq P_{\theta,\omega}[G(Y; \theta, \omega) \leq \alpha] \leq \alpha$$
   (see, for instance, Lange, 2010).
   b) It follows from the above inequality that $\forall \theta \in \Theta_0 \ \forall \omega$
   $$P_{\theta,\omega}[p(X_1, \ldots, X_n; \Theta_0) \leq \alpha] \leq P_{\theta,\omega}[p(X_1, \ldots, X_n; \theta) \leq \alpha] \leq \alpha.$$
   c) Similarly,
   $$P_{\theta,\omega}[\theta \in I_{1-\alpha}(X_1, \ldots, X_n)] = P_{\theta,\omega}[p(X_1, \ldots, X_n; \theta) > \alpha] \geq 1 - \alpha.$$

*Proof of Theorem 2.*
   a) The cumulative distribution function
   $$G(y; \theta, \omega) = P_{\theta,\omega}(Y \leq y)$$
   of the auxiliary statistic $Y = -T_\theta$ leads to
   $$b(x_1, \ldots, x_n; \theta) = \gamma + \sup_{\omega \in S} P_{\theta,\omega}[T_\theta \geq t(x_1, \ldots, x_n; \theta)]$$
   $$\geq \gamma + P_{\theta,\omega}[T_\theta \geq t(x_1, \ldots, x_n; \theta)] = \gamma + G(-t(x_1, \ldots, x_n; \theta); \theta, \omega)$$
   and hence to

$$P_{\theta,\omega}[b(X_1, \ldots, X_n; \theta) \leq \alpha]$$
$$= P_{\theta,\omega}[b(X_1, \ldots, X_n; \theta) \leq \alpha, \omega \in S]$$
$$+ P_{\theta,\omega}[b(X_1, \ldots, X_n; \theta) \leq \alpha, \omega \notin S]$$
$$\leq P_{\theta,\omega}[\gamma + G(Y; \theta, \omega) \leq \alpha] + P_{\theta,\omega}[\omega \notin S]$$
$$\leq \alpha - \gamma + \gamma = \alpha$$

with $S = S(X_1, \ldots, X_n)$.

b) It follows from the above inequality that $\forall \theta \in \Theta_0 \; \forall \omega$
$$P_{\theta,\omega}[b(X_1, \ldots, X_n; \Theta_0) \leq \alpha] \leq P_{\theta,\omega}[b(X_1, \ldots, X_n; \theta) \leq \alpha] \leq \alpha.$$

c) In the same way,
$$P_{\theta,\omega}[\theta \in B_{1-\alpha}(X_1, \ldots, X_n)] = P_{\theta,\omega}[b(X_1, \ldots, X_n; \theta) > \alpha] \geq 1 - \alpha.$$